 \documentclass[%
  aip,jcp,
 amsmath,amssymb,
  reprint,%
 ]{revtex4-1}

\usepackage{graphicx}
\usepackage{dcolumn}
\usepackage{bm}

\usepackage[utf8]{inputenc}
\usepackage[T1]{fontenc}
\usepackage{mathptmx}

\usepackage{epstopdf}
\usepackage[usenames,dvipsnames]{xcolor}
\usepackage[caption=false]{subfig}
\usepackage{booktabs}

\usepackage{soul}
\usepackage{color}

\begin{document}

\preprint{AIP/123-QED}

\title[Modeling elastic spheres]{Modeling of many-body interactions between elastic spheres through symmetry functions}

\author{Emanuele Boattini}
 \email{e.boattini@uu.nl.}
 \affiliation{ 
Soft Condensed Matter, Debye Institute for Nanomaterials Science, Utrecht University, Utrecht, The Netherlands
}
\author{Nina Bezem}
\affiliation{ 
Soft Condensed Matter, Debye Institute for Nanomaterials Science, Utrecht University, Utrecht, The Netherlands
}

\author{Sudeep N. Punnathanam}
\affiliation{ 
Department of Chemical Engineering, Indian Institute of Science, Bangalore 560012, Karnataka, India
}

\author{Frank Smallenburg}
\affiliation{ 
Universit\'e Paris-Saclay, CNRS, Laboratoire de Physique des Solides, 91405 Orsay, France
}
\author{Laura Filion}
\affiliation{ 
Soft Condensed Matter, Debye Institute for Nanomaterials Science, Utrecht University, Utrecht, The Netherlands
}

\date{\today}

\begin{abstract}

Simple models for spherical particles with a soft shell have been shown to self-assemble into numerous crystal phases and even quasicrystals. However, most of these models rely on a simple pairwise interaction, which is usually a valid approximation only in the limit of small deformations, i.e. low densities. In this work, we consider a many-body yet simple model for the evaluation of the elastic energy associated with the deformation of a spherical shell. The resulting energy evaluation, however, is relatively expensive for direct use in simulations. We significantly reduce the associated numerical cost by fitting the potential using a set of symmetry functions. We propose a method for selecting a suitable set of symmetry functions that capture the most relevant features of the particle’s environment in a systematic manner.
The fitted interaction potential is then used in Monte Carlo simulations to draw the phase diagram of the system in two dimensions. The system is found to form both a fluid and a hexagonal crystal phase.

\end{abstract}

\maketitle

\section{Introduction}
Typical spherical colloids with a strong short-ranged repulsion self-assemble into only a few crystalline phases: phase-centered cubic (FCC), hexagonal close-packed (HCP) and body-centered cubic (BCC). In contrast, deformable spherical particles with a “soft” repulsion have been shown to self-assemble into a variety of crystal structures, including open-crystal lattices\cite{watzlawek1999phase}, Frank-Kasper phases\cite{lee2010discovery,gillard2016dodecagonal,hajiw2015evidence}, and even quasicrystals\cite{zeng2004supramolecular}. Because of their rich phase behaviour, these particles are excellent candidates for targeting exotic structures with unique properties.

Even simple models like star polymers\cite{watzlawek1999phase} and Hertzian spheres\cite{pamies2009phase} stabilise numerous crystal phases in three dimensions. The Hertzian potential in particular has attracted significant attention in the context of modeling microgel particles \cite{hashmi2009mechanical, mohanty2014effective, bergman2018new, rovigatti2019connecting, camerin2020microgels}.
In addition to FCC and BCC crystals, stable phases for this model include diamond and body-centered tetragonal (BCT)\cite{pamies2009phase}. 
The phase behaviour of Hertzian spheres is even richer in two dimensions, with a sequence of nine stable phases including a dodecagonal quasicrystal\cite{fomin2018phase}. 
Moreover, tuning the steepness of the interaction further stabilizes an additional octagonal quasicrystal phase, together with various other crystals \cite{zu2017forming}. 


All these models, however, are described by a monotonically-decaying pairwise-interaction, which does not account for possible many-body effects. Furthermore, the Hertz potential describes well the potential energy of two elastic spheres in contact only in the limit of small deformations. Nonetheless, the most interesting behaviour is found at high densities well outside this limit.

An attempt to introduce a more realistic yet simple many-body model for deformable particles was made by Pansu and Sadoc\cite{pansu2017metallurgy}. In their model, particles are treated as soft objects with an internal rigid core and an external soft shell, whose shape depends on the particle’s environment. Specifically, they assume that the particle’s shell fills entirely the space described by its Voronoi cell. The elastic energy is then measured by the shift between the deformed shell and a perfectly spherical corona with a density-dependent size. Note that this model allows also for the stretching of the shell, which results in an attractive force between particles.

Inspired by some of the ideas in their work, here we introduce a many-body model to describe the deformation of spherical particles with a soft shell. In contrast to Ref.\onlinecite{pansu2017metallurgy}, we consider the particle’s shell as a spherical surface of a fixed diameter $\sigma$, i.e. it cannot stretch. Only upon contact of two or more particles do the shells undergo an inward radial deformation. As a result, our model interaction is purely repulsive and accounts only for a possible compression of the particles’ shell. As in Ref. \onlinecite{pansu2017metallurgy}, the surface of the shell is discretized in terms of a large number of small surface elements, which are connected by harmonic springs to the internal core of the particle. With this construction, the energy associated with the shell’s deformation can be expressed as a sum of the energy contributions of the surface elements that are involved in the deformation.

Despite its simplicity, however, the energy evaluation of our model is relatively expensive for direct use in simulations.  Partially due to similar limitations, Ref. \onlinecite{pansu2017metallurgy}  only looked at energetically favoured structures, i.e. the zero temperature limit. A possible solution to this problem is to find a model that can fit very accurately the original interaction at a fraction of the original computational cost.

In recent years, machine learning (ML) techniques have become a powerful tool to approximate complex many-body interactions and predict the properties of molecules and materials based on a few reference calculations \cite{rupp2012fast,faber2016machine,behler2016perspective}. Most of these techniques have been developed to speed up \emph{ab initio} molecular dynamics simulations, where the energy and forces are evaluated with very costly electronic structure methods. However, these methods show significant promise also for complex soft-matter systems.

In order to efficiently interpolate between reference structures, these ML methods usually take as their input some descriptors of the particle's environment rather than the conventional Cartesian coordinates. The role of these descriptors is to encode the relevant physical features of a particle's environment, while satisfying all the symmetries of the problem: invariance with respect to translations, rotations, and permutations of particles of the same species\cite{behler2007generalized,bartok2013representing,glielmo2017accurate, grisafi2018symmetry}. Examples include the Smooth Overlap of Atomic Positions (SOAP) framework\cite{bartok2013representing}, and the symmetry functions (SFs) proposed by Behler and Parrinello \cite{behler2007generalized,behler2011atom}. The former is usually used in combination with Gaussian process regression \cite{bartok2010gaussian,bartok2013machine,musil2018machine}, while the latter have been used in combination with artificial neural networks \cite{behler2007generalized,khaliullin2010graphite,eshet2012microscopic, geiger2013neural, kapil2016high,cheng2016nuclear, singraber2019parallel}.

A crucial step in the optimization of these ML schemes is the selection of a suitable set of descriptors that provide a good balance between efficiency and accuracy. In terms of accuracy, this set must be able to capture all the features of the particle’s environments that are relevant for the prediction of the energy. In terms of efficiency, however, it is desirable to limit the number of descriptors as much as possible. Depending on the family of descriptors and on the regression scheme employed, different selection procedures have been proposed\cite{behler2015constructing,gastegger2018wacsf,imbalzano2018automatic,helfrecht2020structure}. Most of these procedures, however, are designed to work in combination with nonlinear regression schemes, such as artificial neural networks. Only recently, more attention has been given to simpler linear techniques\cite{helfrecht2020structure}.

In this work, we approximate the energy of a particle in our model with a linear combination of the SFs of Behler and Parrinello\cite{behler2007generalized,behler2011atom}. To this end, we introduce a simple and efficient iterative procedure for the selection of a suitable set of SFs. This procedure finds a good balance between computational cost and  accuracy of predictions. Moreover, it provides by definition an excellent indication of whether linear regression or more complex nonlinear schemes are necessary for the problem at hand.

As we will see, by approximating our model interaction as a linear combination of SFs, we are able to speed up the energy evaluation by at least two orders of magnitude. This significant reduction in the computational cost allows us to explore the phase behaviour of the model in two dimensions by means of Monte Carlo simulations.

The remainder of the paper is organised as follows. In Section \ref{sec:model}, we describe our model in detail. In Section \ref{sec:fit}, we first introduce the SFs employed and the fitting procedure, and then discuss the results and the reliability of the fit. In Section \ref{sec:phases} we present the phase behaviour of the model in two dimensions. A final discussion follows in Section \ref{sec:conclusion}.

\section{The model}
\label{sec:model}

\begin{figure}[ht]
\centering
\includegraphics[width=1.\linewidth]{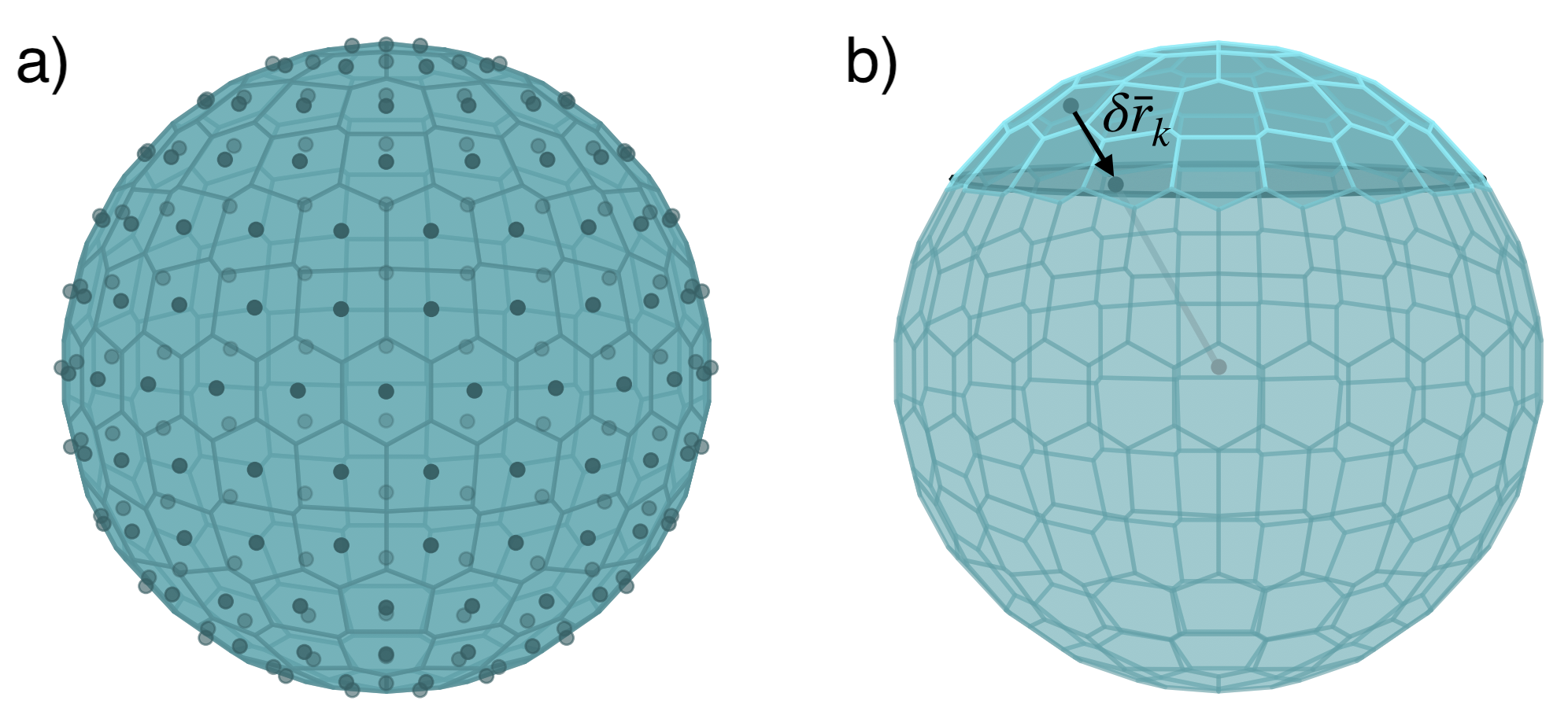}
\caption{(a) Graphical representation of of the surface of a particle's external shell showing the $N_p=200$ points placed on the surface and the associated Voronoi cells. (b) Visualization of the deformation of a portion of the shell due to an overlap with another particle. The deformed portion of the surface and the disk of intersection are highlighted in a darker color. The deformation $\delta\vec{r}_k$ of the $k$th point on the surface is shown explicitly.} 
\label{fig:model}
\end{figure}

We model our deformable colloidal particles as spheres of diameter $\sigma$ consisting of an internal spherical hard-core of diameter $\sigma/2$ and an external deformable shell. When the shells of two particles overlap, they undergo an elastic radial deformation. This deformation causes the regions of the shells involved in the overlap to collapse onto their disk of intersection. In order to model the elastic energy associated with the deformation of the shell, we discretize its surface by decorating it with a large number $N_p$ of approximately equidistributed points. The positions of such points are generated with the algorithm in Ref. \onlinecite{deserno2004generate}.  To each point $k$ on the surface we associate a weight, $w_k$, corresponding to the surface area of its Voronoi cell. This construction is shown in Fig. \ref{fig:model}a.       

\begin{figure}[ht]
\centering
\includegraphics[width=1.0\linewidth]{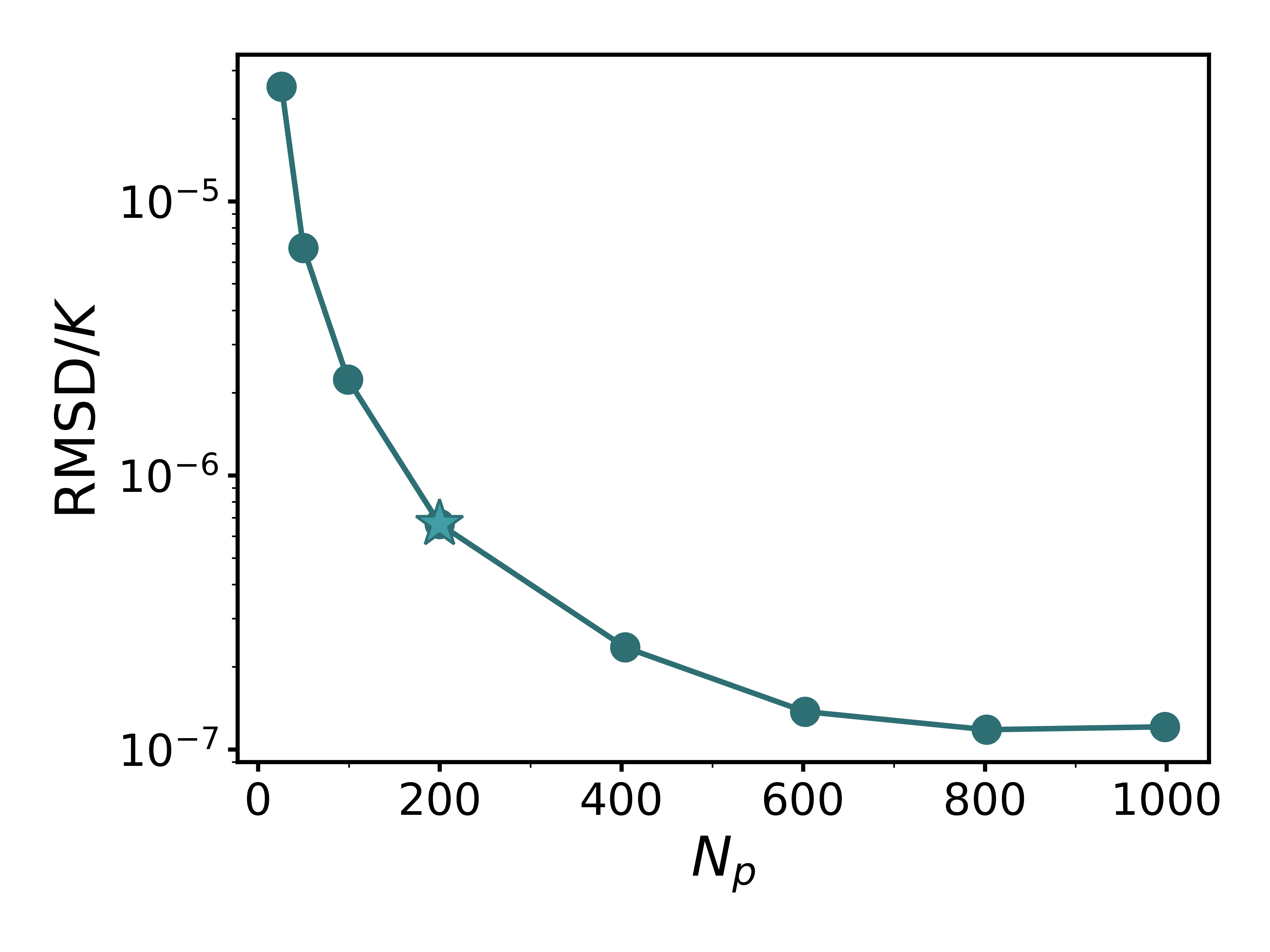}
\caption{Root mean squared deviation (RMSD) from the reference energy as a function of $N_p$. The chosen value, $N_p=200$, is indicated with a star.} 
\label{fig:npoints}
\end{figure}

When a portion of the shell is deformed due to an overlap with another particle, each point on that portion is pushed radially onto the disk of intersection between the two particles. This is shown in Fig. \ref{fig:model}b, where we highlight the part of the surface involved in the deformation with a darker color, and show the explicit deformation $\delta \vec{r}_k$ of the $k$th point on the surface. Given this construction, the elastic energy associated with the shell's deformation of a particle $i$ can be approximated by a weighted sum of the elastic deformation energies associated with each point on its surface,
\begin{equation}
\label{eq:energy}
U_i = Ku_i=\frac{K}{2}\sum_{k=1}^{N_p}\frac{w_k}{\sigma^2}\left(\frac{\delta r_{k}}{\sigma}\right)^2,
\end{equation}
where $\delta r_{k}=\left\lVert\delta \vec{r}_{k}\right\lVert$, and $K$ is a constant with the dimension of an energy. Because of the discretization, the energy in Eq. \ref{eq:energy} is clearly dependent on the value of $N_p$. Moreover, these $N_p$ points are only approximately equidistributed on the surface and their associated Voronoi cells slightly differ both in their shape and surface area (see Fig. \ref{fig:model}a). As a consequence, the energy is also sensitive to the orientation of the particle. One way to alleviate this undesired effect is to estimate the energy as an average over an ensemble of $N_o$ randomly chosen orientations. Here, we set $N_o=100$, while we optimize the value of $N_p$ by checking the convergence of the energy as $N_p$ goes to infinity. In the simple case of two overlapping particles $i$ and $j$, the interaction potential in the limit of small surface elements can be evaluated analytically via simple integration, and reads
\begin{equation}
\label{eq:analytical}
U_i(r_{ij}) = \frac{K\pi}{16}\left[1-\left(\frac{r_{ij}}{\sigma}\right)^2 + 2\left(\frac{r_{ij}}{\sigma}\right)\log \left(\frac{r_{ij}}{\sigma}\right)\right],
\end{equation}
where $r_{ij}$ is the distance between the two particles.
We use this expression as a reference and evaluate the error introduced by the discretization as a function of $N_p$. This is done by evaluating the energy in Eq. \ref{eq:energy} for several equally spaced distances in the interval $r_{ij}/\sigma \in(0.5,1)$ and for different values of $N_p$. The error is then evaluated as the root mean squared deviation (RMSD) from the reference energy. As shown in Fig. \ref{fig:npoints}, this deviation becomes small as $N_p$ increases and reaches a plateau at about $N_p\approx600$. We attribute this plateau to the numerical precision of our calculations. As a reasonable compromise between accuracy and efficiency, we choose a value of $N_p=200$.

\begin{figure}[ht]
\centering
\includegraphics[width=1.\linewidth]{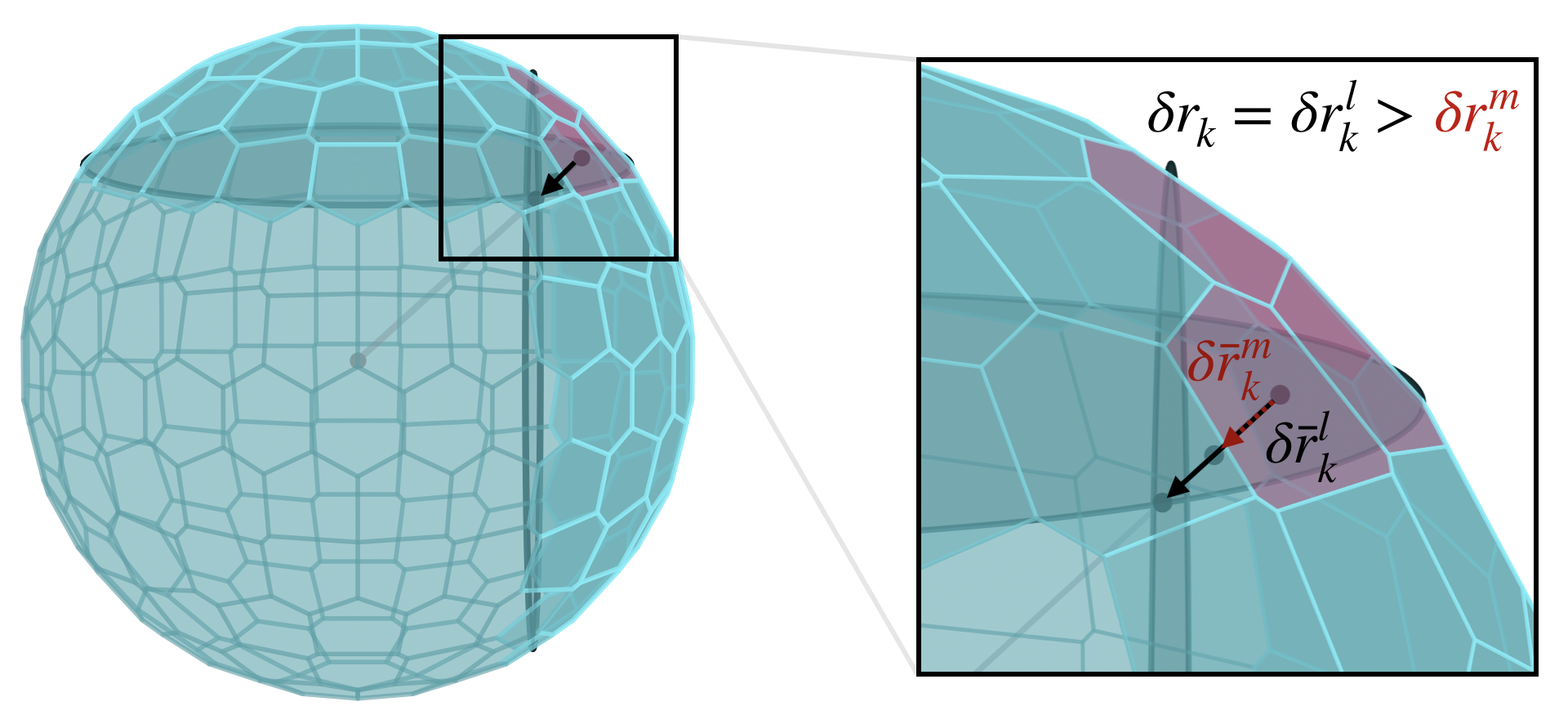}
\caption{Visualization of the shell's deformation of a particle overlapping with two other particles. Highlighted in red is the region where the many-body effect of the interaction expressed by Eq. \ref{eq:deformation} comes into play.} 
\label{fig:manybody}
\end{figure}

The elastic energy introduced in Eq. \ref{eq:energy} in the case of two overlapping particles can be extended easily to the case of more interacting particles by taking particular care in the evaluation of the deformation $\delta \vec{r}_k$ of the points on the surface. When a particle $i$ is overlapping with more neighboring particles, one can repeat the same construction presented in Fig. \ref{fig:model}b for each of the overlapping particles, and compute, for every point $k$ on its surface, the deformation due to each of these particles. Then, the actual deformation of point $k$ is simply given by
\begin{equation}
\label{eq:deformation}
\delta r_k = \max_{j} \delta r^j_k,
\end{equation}
where $\delta r^j_k$ is the deformation of point $k$ due to the presence of particle $j$ alone, and $j$ runs over all the neighboring particles overlapping particle $i$ in the region containing point $k$. An example of such a construction in the case of three overlapping particles is shown in Fig. \ref{fig:manybody}.  

\begin{figure*}[ht]
\centering
\includegraphics[width=1.0\linewidth]{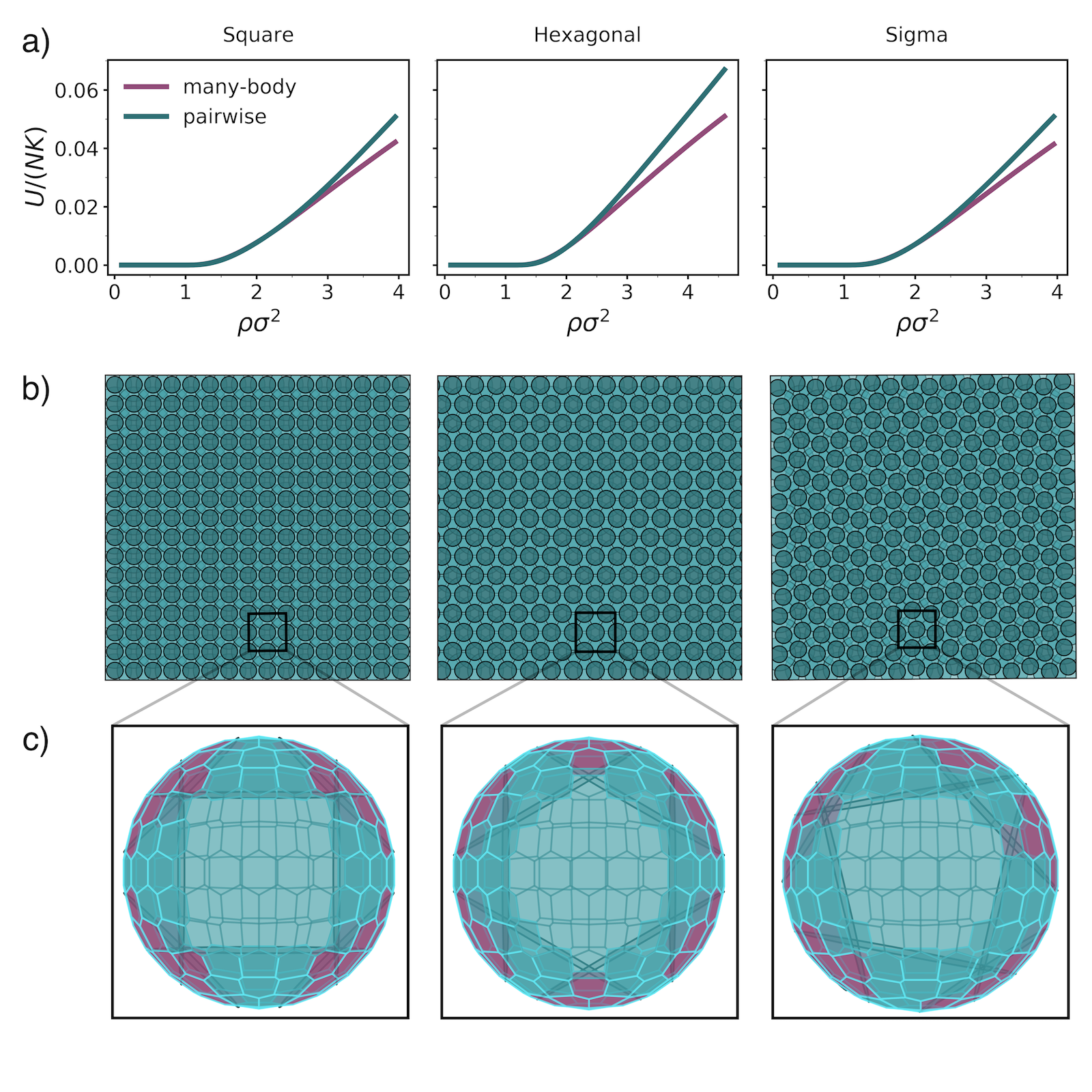}
\caption{(a) Energy per particle in the square (left), hexagonal (center) and sigma (right) phases as a function of density. The energy is evaluated both in a many-body and pairwise fashion by considering or not the correction introduced in Eq. \ref{eq:deformation}. (b) Snapshots of a typical configuration of the three phases. (c) Typical shell's deformation of a particle in these phases. The deformed portion of the shell is highlighted with a darker color. The red color indicates the regions where the many-body effect comes in to play.}
\label{fig:encrys}
\end{figure*}

Taking the maximum in Eq. \ref{eq:deformation} allows us to correctly evaluate the actual deformation of a surface point and avoids overestimating the associated elastic energy. One could think of the effect of Eq. \ref{eq:deformation} as a many-body correction on top of a pairwise interaction. In a pairwise fashion, the deformation energy of a particle is given by the sum of the deformation energies caused independently by each of its overlapping neighbours, i.e. allowing the overestimation of the energy for some of the surface points. With such a picture in mind, the many-body effect introduced in this model always has a negative sign compared to the pairwise interaction. To further stress this point, we evaluate the energy per particle in perfect configurations of the square, hexagonal and sigma phases both in a pairwise fashion, i.e. without the correction introduced by Eq. \ref{eq:deformation}, and in a many-body fashion, i.e. by considering the actual deformation of the surface points. A comparison of these energies as a function of density is shown in Fig. \ref{fig:encrys}a. Figure  \ref{fig:encrys}b shows a typical configuration of the three phases and  Fig. \ref{fig:encrys}c shows the typical shell deformation of a particle in these phases. 

In the present work, we aim to study the phase behaviour of these particles in two dimensions by means of Monte Carlo (MC) simulations. However, the large number of points required for the discretization of the particle's shell and the need of averaging over different orientations make the evaluation of the interaction computationally very demanding. This high computational cost limits the size of the system and the time scales that can be assessed in simulations.
We overcome these difficulties by fitting the interaction as a linear combination of the SFs introduced by Behler and Parrinello \cite{behler2007generalized}. As we will see, the fit speeds up the energy evaluation of at least two orders of magnitude, making it possible to perform long simulations of large systems.

In the next section, we present the details of the fitting procedure and discuss the results.

\section{Fitting Procedure}
\label{sec:fit}
In the following, we focus on the SFs of Behler and Parrinello and use them in combination with simple linear regression for approximating the energy of a particle in our model. Specifically, we first describe how we generate representative configurations of the particles' configurations, i.e. the training data set. Then, we introduce the types of SFs used in this work and the fitting procedure. Finally, we discuss the accuracy and reliability of the fit.

\subsection{Training data set}
\label{sec:trainingset}

As we are interested in studying the phase behaviour of this model in two dimensions, the training data set should include representative examples of the particle's typical environments for a wide range of different densities and interaction strengths $K$. We perform MC simulations of the model in the canonical ensemble, i.e. constant number of particles $N$, area $A$, and temperature $T$. Note that, since the averaging over different orientations introduces statistical uncertainty in the energy evaluation, in these simulations we use the penalty method introduced in Ref. \onlinecite{ceperley1999penalty}. In this method, the standard acceptance rule proposed by Metropolis \emph{et al.}\cite{metropolis1953equation} is modified by adding to the energy difference an error-dependent penalty term. The effect of this term is to correct, on average, for the presence of the noise.

To build the training data set, we consider a relatively small system size of $N=64$ particles and restrict our attention to densities for which many-body effects are present. Specifically, simulations are performed for several densities in the range $\rho\sigma^2\in[2.4,4.6]$ starting from a hexagonal crystal configuration, and $\rho\sigma^2\in[2.4,4]$ starting from a square crystal configuration. In both cases, the density spacing considered is $\delta\rho\sigma^2=0.025$. Moreover, for each density and initial configuration, we consider different values of the interaction strength: $\beta K\in\{1,10,50,100,1000\}$, where $\beta = 1/k_B T$ with $k_B$ Boltzmann's constant and $T$ the temperature. 

From each simulation, we save $50$ independent snapshots together with the corresponding particles' dimensionless energies, $u_i=U_i/K$. Finally, 10 particles from each snapshot are randomly selected and included in our data set. $80\%$ of this data set is used for the training, while the remaining $20\%$ is used for testing the model.

\subsection{Symmetry functions}
To describe the local environment of a particle we use the SFs introduced by Behler and Parrinello for constructing high-dimensional neural network potentials \cite{behler2007generalized}. These SFs are described in great detail in Refs. \onlinecite{behler2011atom,behler2015constructing} and have been used as inputs for atomic feed-forward neural networks in order to provide the atomic energy contributions of different materials and molecules \cite{behler2007generalized,khaliullin2010graphite,eshet2012microscopic,kapil2016high,cheng2016nuclear}. Here, we briefly describe the form of the two types of SFs employed in this work.

The first type of SFs, $G_2$, provides information on the pair correlations between the reference particle and its neighbours, i.e. all particles closer than a fixed cutoff distance $r_c$ to the reference particle. For a given particle $i$, $G^2_i$ is defined as
\begin{equation}
\label{eq:g2}
G^2_i = \sum_{j}^{} e^{\eta(r_{ij}-R_s)^2}  f_c(r_{ij}),
\end{equation}
where $r_{ij}$ is the distance between particles $i$ and $j$, $\eta$ and $R_s$ are two parameters that control the width and the position of the Gaussian with respect to particle $i$, and the sum runs over all neighbours $j$ being closer than $r_c$. Additionally, $f_c(r_{ij})$ is a cutoff function: a monotonically decreasing function that smoothly goes to 0 in both value and slope at the cutoff distance $r_c$. Here, we consider a cutoff function of the form
\begin{equation}
\label{eq:fc}
f_c(r_{ij}) = 
\begin{cases}
0.5\left[\cos\left(\pi r_{ij}/{r_c}\right)+1\right] & \text{for } r_{ij}\le r_c\\
0 & \text{for } r_{ij} > r_c.
\end{cases}
\end{equation}

The second type of SFs, $G_3$, provides information on angular correlations and it is defined as
\begin{equation}
\label{eq:g3}
\begin{split}
G^3_i =& 2^{1-\xi}\sum_{j,k\ne i}^{} (1+\lambda \cos\theta_{ijk})^\xi e^{\eta(r^2_{ij}+r^2_{ik}+r^2_{jk})} \times\\ & f_c(r_{ij}) f_c(r_{ik}) f_c(r_{jk}),
\end{split}
\end{equation}
where the indices $j$ and $k$ run over all the neighbours of particle $i$, and $\xi$, $\eta$, and $\lambda$ are three parameters that determine the shape of the function. The parameter $\lambda$ can have the values $+1$ or $-1$ and determines the angle $\theta_{ijk}$ at which the angular part of the function has its maximum ($\theta_{ijk}=0^o$ for $\lambda=1$, and $\theta_{ijk}=180^o$ for $\lambda=-1$). The angular resolution is provided by the parameter $\xi$, while $\eta$ controls the radial resolution.

Here, the goal is to express the particle's deformation energy, $u_i=U_i/K$, as a function of a suitable set of SFs. The selection of this set, i.e. the number of SFs to use and their parameters, is arguably the most crucial step in the optimization of the fitting procedure. In the context of neural-network-based potentials, this is usually done by evaluating empirically how the accuracy of the model depends on the set of SFs employed. Based on this idea, one can adopt several optimization strategies in order to find a proper set of SFs. For instance, a recent work proposed the use of genetic algorithms as a method for an optimal selection\cite{gastegger2018wacsf}. The main drawback of such a procedure is that the training of the model has to be repeated for every considered set.

A completely different approach for an efficient and automatic selection was proposed in Ref. \onlinecite{imbalzano2018automatic}. The various methods introduced there are based solely on the knowledge of the geometry of the particles' environments, and do not rely on the energy, nor on the performance of the model that results from a given choice of the SFs. Instead, the common idea behind these methods is to choose SFs which are as diverse as possible by e.g. minimizing their linear correlation. This avoids including redundant information and allows one to capture different aspects of the particle's environment using a relatively small set of SFs. One possible risk of this approach in terms of efficiency is the inclusion of SFs which poorly correlate with the energy. These SFs would be sensitive to aspects of the particle's environment that hardly influence the particle's energy. While their inclusion would not harm the accuracy of the model, their evaluation would constitute an unnecessary numerical overhead.

In the following, we draw inspiration from Ref. \onlinecite{imbalzano2018automatic} and introduce a new efficient and automatic procedure for the selection of SFs. As we will see, the proposed method is by definition ideally suited to work in combination with simple linear regression, but could also be used as the basis of a nonlinear regression scheme.

\subsection{Selection of SFs}
\label{sec:selection}

As in Ref. \onlinecite{imbalzano2018automatic}, the first step of the procedure involves the creation of a large but manageable pool of candidate SFs. This is done by calculating, for every particle in the data set, several SFs with different sets of parameters. At this stage, the parameters are chosen following simple heuristic rules with the goal of capturing most of the possible correlations within the cutoff radius. Here, we fix the cutoff to the range of the interaction defined in our model, i.e. $r_c=\sigma$.

We build a set of radial symmetry functions $G_2$ consisting of Gaussian functions (Eq. \ref{eq:g2}) both centered around the reference particle ($R_s = 0$) and displaced away from the reference particle ($R_s > 0$). For the SFs with $R_s = 0$, we consider 15 different Gaussian widths in the range $\eta\in[0.001, 24]$. For the other radial symmetry functions, we vary the center of the Gaussian among 9 equally spaced values in the range $R_s/r_c\in[0.1, 0.9]$, with Gaussian widths chosen on a logarithmic scale: ${\eta}\in\{1,2,4,8,16\}$.

The $G_3$ angular SFs are generated by setting $\lambda\in\{-1,1\}$, $\eta\in \{0.01,0.1,1,2,4,8\}$, and $\xi\in \{1,2,4,8\}$.

With these choices, our pool of candidates consists of $M=108$ SFs. One could in principle consider a larger pool, making the description of the environment more complete at the expense, however, of a greater computational cost. In this work, we found this pool to be sufficiently large.


The second step of the procedure consists of selecting from the initial pool a subset of $N_s < M$ SFs that captures the most relevant features of the particle's environment and can be used as the basis of a regression scheme to approximate the particle's energy. In this step, the SFs are selected from the pool one after the other in a way that maximizes the overall correlation with the target energy.

The first SF that is selected is the one with the highest linear correlation with the energy, i.e. the one that alone best predicts the energy. As a measure of the correlation, we use the square of the Pearson correlation coefficient, defined as
\begin{equation}
\label{eq:pearson}
c_k = \frac
{\sum_{i}^{}\left(S_k(i)-\bar{S}_k\right) \left(u_i-\bar{u}\right)}
{\sigma_{SD}(S_k)\sigma_{SD}(u)},
\end{equation}
where $S_k(i)$ and $u_i$ are, respectively, the $k$-th SF in the pool and the energy of the $i$-th particle in the data set, $\bar{S}_k$ and $\bar{u}$ are their arithmetic means evaluated over the whole data set, and $\sigma_{SD}(S_k)$ and $\sigma_{SD}(u)$ are their standard deviations. Then, the second SF selected is chosen to be the one that maximizes the increase in the correlation with the energy. The linear correlation between a set of SFs and the energy is quantified by the coefficient of multiple correlation, $R$, whose square is given by
\begin{equation}
\label{eq:K2}
R^2 = \mathbf{c}^\text{T}\mathbf{R}^{-1}\mathbf{c}. 
\end{equation}
Here, $\mathbf{c}^\text{T}=(c_1, c_2,\dots)$ is the vector whose $i$-th component is the Pearson correlation, $c_i$, between the $i$-th SF in the set and the energy, while $\mathbf{R}$ is the correlation matrix of the current set of SFs. Specifically, the element $\text{R}_{ij}$ of this matrix is the Pearson correlation between the $i$-th and the $j$-th SFs in the set. Note that in the case of only one SF $S_i$, $R^2$ reduces to $c_i^2$. $R^2$ can also be computed as the fraction of variance that is explained by a linear fit (including an intercept) of the energy in terms of the SFs in the set. Although computationally slightly more expensive than the evaluation of Eq.\ref{eq:K2}, we found this second method to be numerically more stable.

By maximizing the increase in the correlation with the energy, we aim to select new SFs that add relevant information to the currently selected set, while penalizing both (i) highly correlated SFs with only redundant information, and (ii) SFs which are sensitive to aspects of the particle’s environment that poorly correlate with the particle's energy. The inclusion of point (ii) in our selection procedure is arguably the main difference with the methods proposed in Ref. \onlinecite{imbalzano2018automatic}, where only point (i) is addressed. The described process can be repeated iteratively in order to select new SFs until $R^2$ stops increasing appreciably. This, indeed, indicates that the remaining SFs in the pool add negligible information, and gives us a simple rule to optimize the number of selected SFs.


Another advantage of our procedure is that the value of $R^2$ is by definition a quantitative measure of how well a linear combination of the selected SFs can approximate the particle's energy. As a result, its final value represents an excellent indication of whether linear regression or more complex nonlinear schemes are necessary for the problem at hand. When sufficient, using simple linear regression instead of nonlinear neural networks, for instance, might have some important advantages:
(i) the parameters' optimization is deterministic instead of stochastic, which makes the fitting process, or training, much simpler in terms of efficiency and accuracy;
(ii) the risk of overfitting is considerably lower;
(iii) the resulting model is cheaper in terms of computational cost.

In the next section, we present the results of the fit and compare its efficiency and reliability with the original model.

\subsection{Fit results and validation}
\label{sec:validation}
\begin{figure}[ht]
\centering
\includegraphics[width=1.0\linewidth]{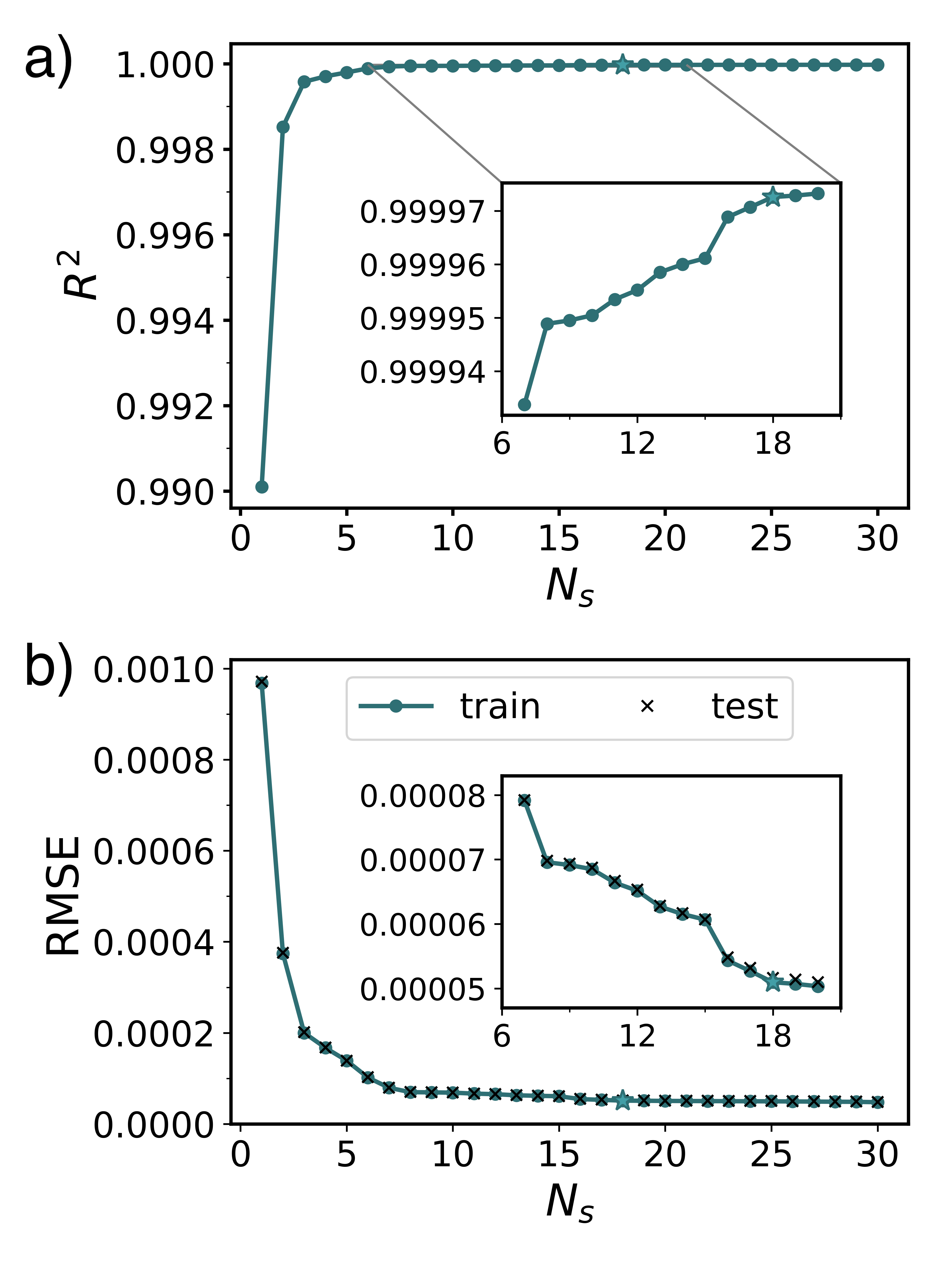}
\caption{(a) Square of the correlation coefficient, ${R}^2$, and (b) root mean squared error (RMSE) as a function of the number of symmetry functions employed, $N_s$. The RMSE is shown for both the training and the test sets. The chosen value, $N_s=18$, is indicated with a star.} 
\label{fig:corr}
\end{figure}

We now present the results of the selection procedure in our problem. Fig. \ref{fig:corr} shows (a) $R^2$ and (b) the root mean squared error (RMSE) of the corresponding linear fit as a function of the number of selected SFs. The RMSE is shown both for the training and test sets. Note that $R^2$ and the RMSE are related by a simple transformation, i.e. $R^2=1-\text{RMSE}^2/\sigma_{SD}^2(u)$.

The first thing to notice is that with only a few SFs $R^2$ approaches very closely its maximum possible value, i.e. $R^2=1$. This clearly indicates that even linear regression using a low number of SFs can very accurately approximate our target interaction.

Next, if we look closely at the variation of the RMSE as a function of $N_s$ (Fig.\ref{fig:corr}b), we can identify three distinct regimes. Initially, the error decreases very rapidly: by about one order of magnitude going from $N_s=1$ to $N_s=7$. After that, the error keeps decreasing more slowly until $N_s\simeq18$. Finally, after $N_s=18$, the error essentially stays constant. From this picture, we can distinguish two obvious choices for the number of symmetry functions:  $N_s = 7$ and $N_s = 18$. Both will provide a good balance between accuracy and efficiency, but clearly choosing $N_s = 7$ leads to a more efficient model. Limiting the number of selected SFs is particularly important when using them as the input of a neural network. In such cases, the input size influences also the size of the following layers in the network and, as a result, has a stronger impact on the overall efficiency of the model. However, as we are using linear regression, we opted instead for a slightly less efficient but more accurate choice of $N_s=18$.

\begin{figure}[ht]
\centering
\includegraphics[width=1.0\linewidth]{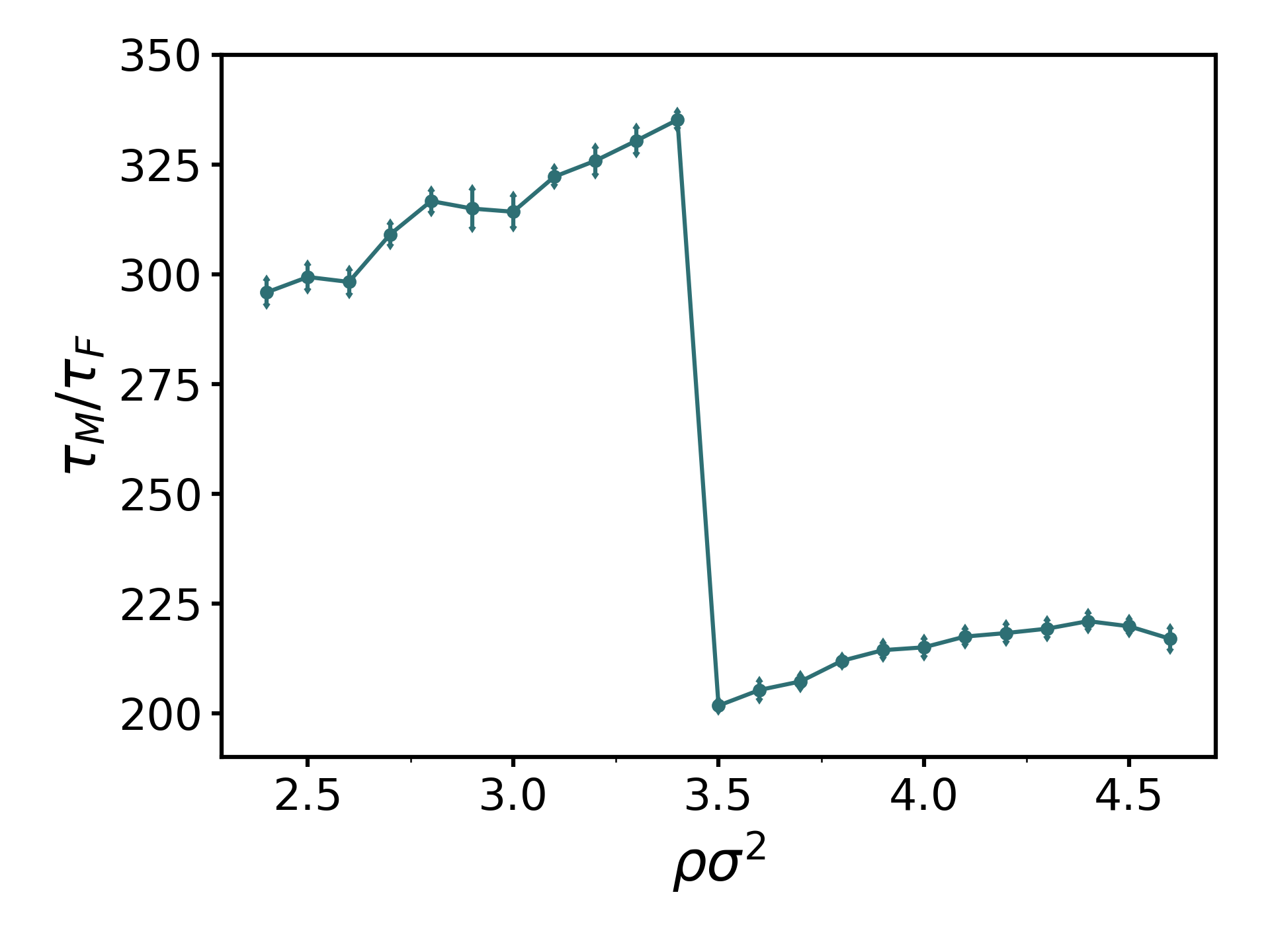}
\caption{Ratio between the computational times required for the energy evaluation in a hexagonal crystal using the original model, $\tau_M$, and the fitted potential, $\tau_F$, as a function of density.} 
\label{fig:speedup}
\end{figure}

To quantify the speed-up achieved with the fit, we compared the computational times required for the energy evaluation using the original model ($\tau_M$) and the fitted potential ($\tau_F$). Since these times depend differently on the density and on the number of neighbours interacting with each particle, we considered a hexagonal crystal at different densities. The ratio $\tau_M/\tau_F$, represented in Fig.\ref{fig:speedup}, clearly shows that the fit speeds up the energy evaluation by at least two orders of magnitude at all densities. A discontinuous jump is observed at a density of $\rho\sigma^2=3.5$, where an extra shell of neighbors enters the interaction range of any given particle. Although these extra neighbors usually do not contribute to the energy, our Monte Carlo simulation still takes them into account as possibly interacting neighbors, and hence around this density, the computation cost goes up significantly. Nonetheless, the fit is at least 200 times faster than the original model even at high densities.

We now turn the discussion to the reliability of our fitted potential. A first simple indication of the general accuracy of the fit is given by its performance on data that was not used in the training. Fig.\ref{fig:corr}b shows that the RMSE on the training and test sets are approximately equal for every value of $N_s$, indicating that the fit generalizes well to "previously unseen'' data. Although this is a good measure of the quality of the fit, it is also important to verify its reliability when computing other properties of the system. To this end, we performed MC simulations in the isothermal-isobaric ensemble ($NPT$) with the original model and with the fitted potential. For several pressures and different values of the interaction strength $\beta K$, we compare the estimated equilibrium density of the system. Note that we considered a system size of $N=64$ particles and started all simulations from a high density hexagonal crystal. Furthermore, to speed up the simulations using the original model, we limited the number of orientations in the averaging to $N_o=10$.

\begin{figure*}[ht]
\centering
\includegraphics[width=1.0\linewidth]{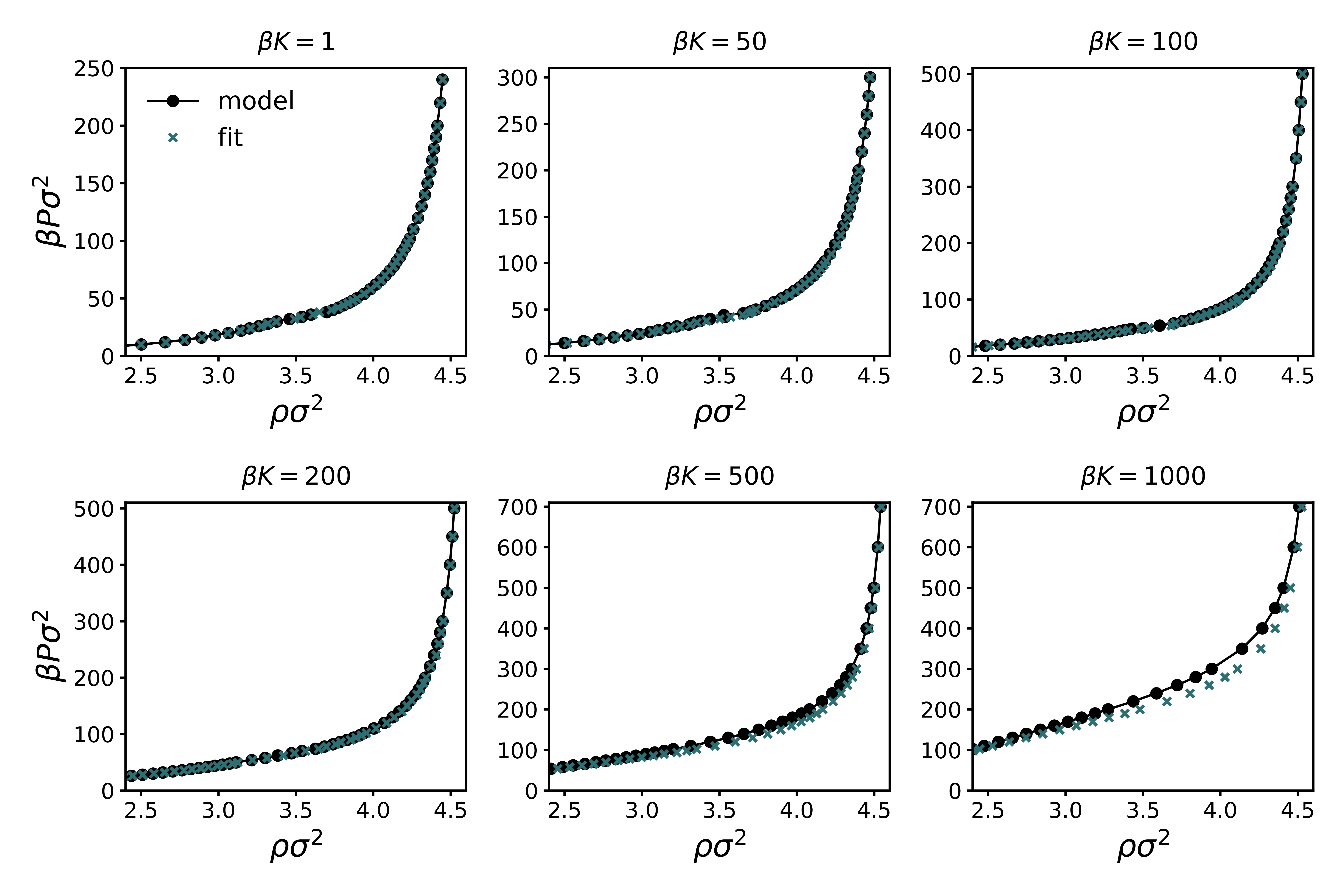}
\caption{Comparison of the pressure as a function of density obtained with the original model and the fitted potential for $\beta K=1$ (top left), $\beta K=50$ (top center), $\beta K=100$ (top right), $\beta K=200$ (bottom left), $\beta K=500$ (bottom center), and $\beta K=1000$ (bottom right). Simulations were performed with a system size of $N=64$ particles and starting from a hexagonal crystal configuration at a density $\rho\sigma^2=4.6$.} 
\label{fig:validation}
\end{figure*}

Fig.\ref{fig:validation} shows a comparison of the results obtained with the model and with the fit for six values of $\beta K$. Up to a value of $\beta K=200$, we find an excellent agreement between the two models. The only appreciable differences are observed for a few pressures in the vicinity of a first-order phase transition. At those pressures, the system repeatedly jumped from one phase to the other during the simulations. As each jump requires the system to overcome a free energy barrier, the time spent in the two phases depends on the particular run and affects the final estimate of the density. As a consequence, similar differences would be observed even if running distinct simulations with the same model.

For larger values of $\beta K$, however, discrepancies between the original model and the fit start appearing also at points far from the phase transition. This can be seen in Fig.\ref{fig:validation} for $\beta K=500$, and especially for $\beta K=1000$, where the difference becomes more pronounced. There is a simple explanation to this. For high values of $\beta K$, the error in the fit (which scales with the prefactor $K$) becomes comparable to the typical energy fluctuations (i.e. comparable to $k_B T$). This leads to observable errors in the overall behavior of the system, and hence in the equation of state. In our case, this clearly starts happening for an interaction strength of at least $\beta K=500$, after which our fitted potential becomes less reliable.

\section{Phase behaviour}
\label{sec:phases}

We investigated the phase behaviour of the system in two dimensions by performing $NPT$-MC simulations with the fitted interaction potential. Specifically, we determined the equation of state (EOS) of the system, i.e. the pressure as a function of the equilibrium density, for several values of the interaction strength in the range $\beta K\in[1,500]$. We repeated this analysis for different initial conditions of the system: a low-density fluid phase, and high-density hexagonal, square and sigma phases. We chose these initial configurations as similar phases have been observed in similar models for deformable colloids. In all cases we considered a system size of $N=256$ particles. Additionally, during these simulations, we let the two axes of the box change independently in order to let the box shape adapt.

To assess the structure and discriminate between different phases, we computed the averaged $m$-fold bond orientational order parameter $\chi_m$, defined as\cite{weber1995melting} 
\begin{equation}
\label{boop}
\chi_m=\left<\left|\frac{1}{N_b(j)} \sum_{k=1}^{N_b(j)} \exp(im\theta_{\mathbf{r}_{jk}})\right|^2\right>.
\end{equation}
Here, $m$ is an integer associated with the symmetry of interest, $\mathbf{r}_{jk}$ is the vector connecting particles $j$ and $k$, and $\theta_{\mathbf{r}_{jk}}$ is the angle between $\mathbf{r}_{jk}$ and an arbitrary axis. Additionally, the sum over $k$ runs over the $N_b(j)$ nearest neighbours of particle $j$. The set of nearest neighbors of each particle is identified
with a two-dimensional adaptation of the parameter-free criterion called SANN (solid angle nearest neighbor)\cite{van2012parameter}.

\begin{figure*}[ht]
\centering
\includegraphics[width=1.0\linewidth]{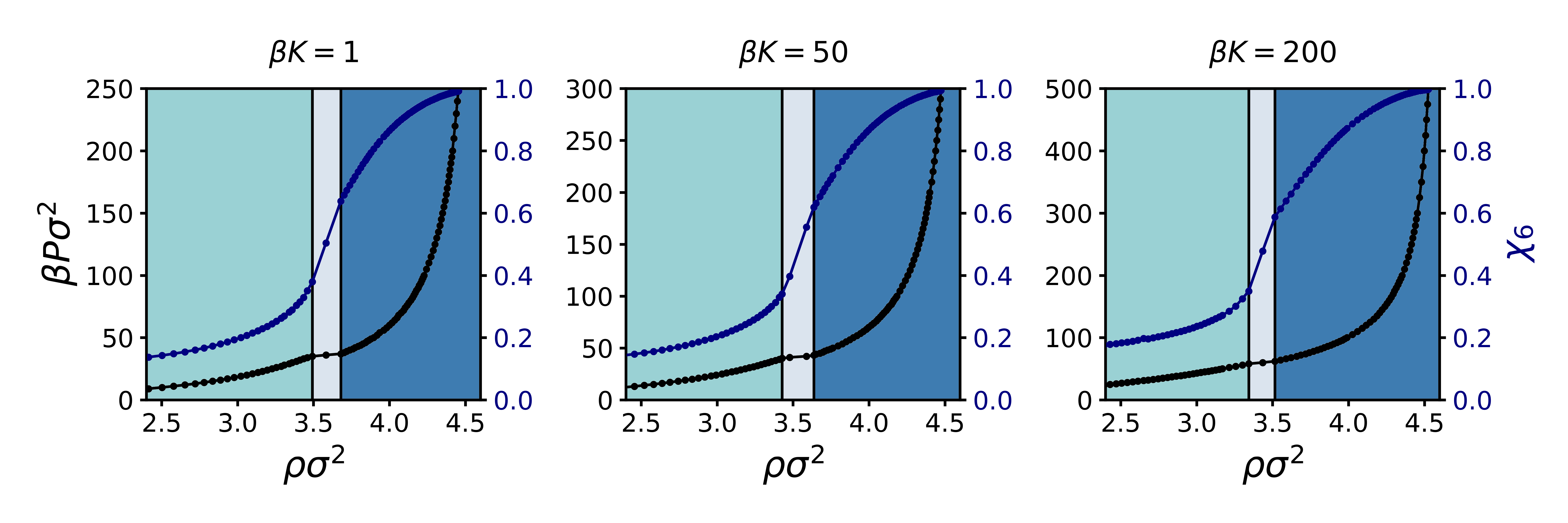}
\caption{Pressure (black) and $\chi_6$ (blue) as a function of density for $\beta K=1$ (left), $\beta K=50$ (center) and $\beta K=200$ (right).} 
\label{fig:eos}
\end{figure*}

For all values of $\beta K$ and initial conditions considered, the system stabilized a fluid phase at low densities and a hexagonal crystal phase at higher densities. In Fig.\ref{fig:eos}, we show the measured EOS and the order parameter $\chi_6$ for three values of $\beta K$. $\chi_6$ measures the degree of $6$-fold symmetry in the system and it is expected to be high in the hexagonal phase. The background colors in Fig.\ref{fig:eos} indicate the regions of stability of the two phases, while the color grey indicates the coexistence region. Points falling in this region correspond to pressures at which the system jumped multiple times from one phase to the other during the simulation. 

\begin{figure}[ht]
\centering
\includegraphics[width=1.0\linewidth]{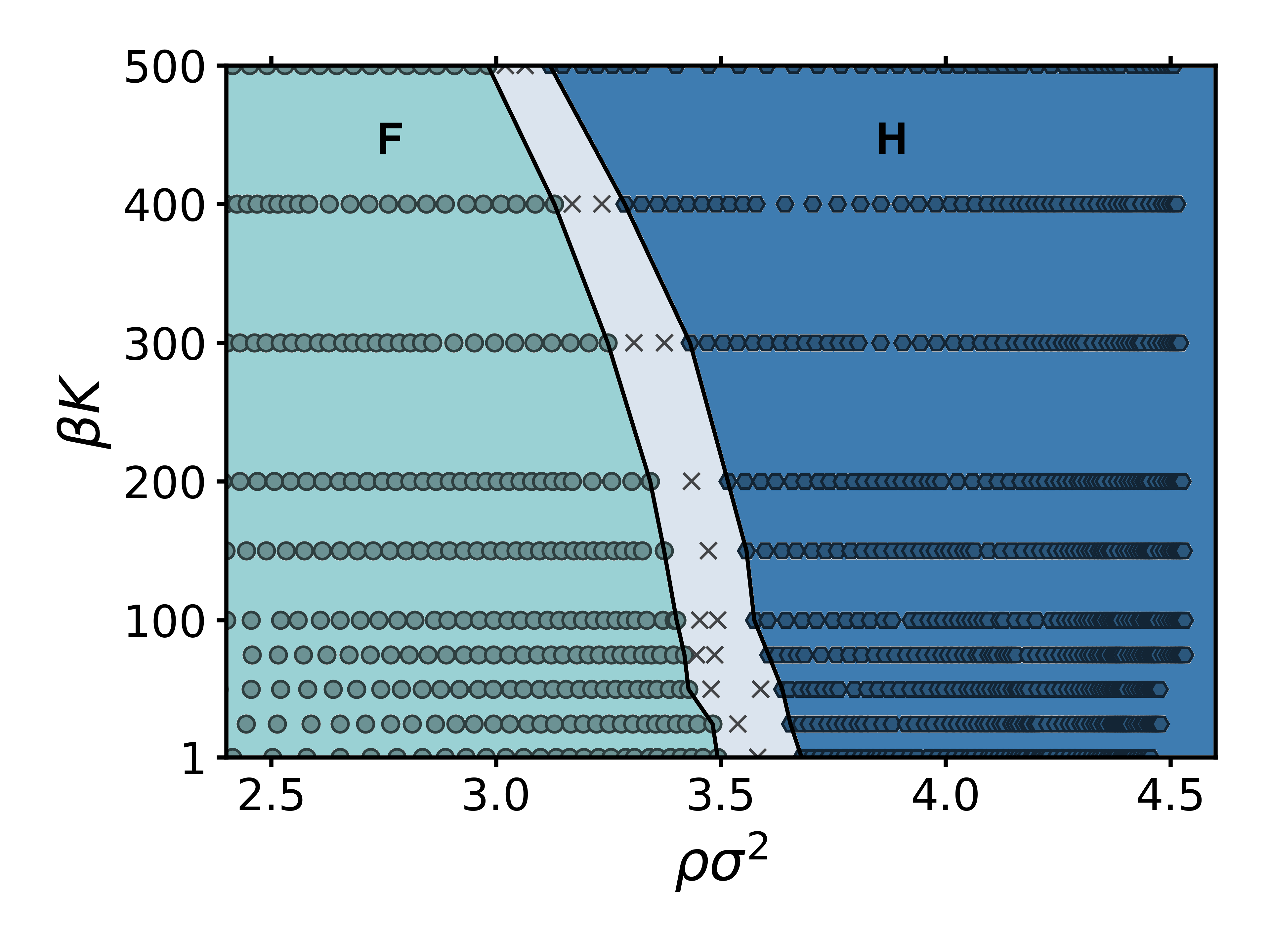}
\caption{Phase diagram as a function of the interaction strength, $\beta K$, and the density, $\rho\sigma^2$. The stable phases are the fluid phase (F) and hexagonal crystal phase (H). The grey area denotes the coexistence region between the two phases.} 
\label{fig:phases}
\end{figure}

Finally, from these EOS we determined the phase behaviour of the system as a function of the interaction strength and the density. The constructed phase diagram is shown in Fig.\ref{fig:phases}. Note that this phase diagram is actually much simpler than that of the analytic, 2-body Hertzian potential, which displays a variety of complex phases \cite{pamies2009phase}.

\section{Conclusions}
\label{sec:conclusion}

In short, we have introduced a simple and efficient method to fit computationally expensive many-body interactions using linear combinations of symmetry functions. In particular, our approach selects an effective set of symmetry functions with an iterative procedure performed on a representative set of sample configurations. We used this approach to fit the interaction potential of colloids coated with a deformable shell, speeding up the energy evaluation by at least to orders of magnitude. Using Monte Carlo simulations of the fitted potential, we scan the phase diagram of the colloidal model system over a range of temperatures and densities, revealing a fluid and a single stable hexagonal crystal phase. 

While the model we investigated is relatively simple, the computational speedup demonstrates that our approach provides an effective way forward in situations where interaction potentials between particles are too computationally expensive to be tractable in standard computer simulations. This not only applies to simulating deformable colloids, but also to {\it ab initio} simulations of atomic systems, where interactions between atoms are complicated by quantum effects. Similarly, the same approach can straightforwardly be extended to fitting {\em effective} interactions between particles, e.g. in systems containing polymer chains (in solution or grafted onto particles) which carry a large number of degrees of freedom.

The main strengths of the fitting approach we propose here are its efficiency and its flexibility. The efficiency stems from the fact that the iterative method used to select symmetry functions is directly aimed at fitting the desired energy function, and thus avoids including SFs that do not correlate with the target function. This criterion relies on the fact that our final fitting scheme is a simple linear fit, which ensures that the expected impact of adding a new SF can be gauged based on its linear correlation with the energy function and the previous set of SFs. The flexibility stems from the fact that the method does not rely on a specific choice for the pool of SFs we select from. Here, we only include radial and angular functions, since we know from physical considerations that the particles are isotropic. However, including SFs that take into account the orientation of the particles is straightforward. Hence, we believe that the proposed approach will be a valuable tool for speeding up the simulation of particles with complex interactions.

\begin{acknowledgments}
We would like to thank Gabriele Maria Coli and Massimiliano Chiappini for many useful discussions.
L. F., E. B., and S. P. gratefully acknowledge funding from the Netherlands Organisation for Scientific Research (NWO) [grant number 16DDS004]. F. S. gratefully acknowledges funding from Agence National de la Recherche [grant number ANR-18-CE09-00].
\end{acknowledgments}

\section*{Data availability}
The data that support the findings of this study are available from the corresponding author upon reasonable request.

\bibliography{ref}

\end{document}